\documentclass[openany]{nature}
\usepackage{graphicx}
\usepackage[none]{hyphenat}
\usepackage{tikz}
\usepackage{ulem} 
\usepackage{lineno}
\usepackage{amssymb}
\usepackage{amsmath}
\usepackage{caption}
\usepackage{subcaption}
\usepackage{tabularx}
\usepackage{booktabs,multirow}
\usepackage{multicol}
\usepackage{threeparttable}

\makeatletter

\let\saved@includegraphics\includegraphics
\AtBeginDocument{\let\includegraphics\saved@includegraphics}
\renewenvironment*{figure}{\@float{figure}}{\end@float}
\makeatother

\title{An Energetic Hot Wind from the Low-luminosity Active Galactic Nucleus M81*}

\author{Fangzheng Shi$^{1,2}$, Zhiyuan Li$^{1,2}$, Feng Yuan$^{3,4}$ \& Bocheng Zhu$^{3,4}$}

\begin{document}
\maketitle

\begin{affiliations}
 \item School of Astronomy and Space Science, Nanjing University, Nanjing 210023, China
 \item Key Laboratory of Modern Astronomy and Astrophysics, Nanjing University, Nanjing 210023, China
 \item Shanghai Astronomical Observatory, Chinese Academy of Sciences, Shanghai 200030, China
 \item University of Chinese Academy of Sciences, 19A Yuquan Road, Beijing 100049, China
\end{affiliations}

\begin{abstract}
For most of their lifetime, super-massive black holes (SMBHs) commonly found in galactic nuclei obtain mass from the ambient at a rate well below the Eddington limit\cite{2008ARA&A..46..475H}, which is mediated by a radiatively inefficient, hot accretion flow\cite{2014ARA&A..52..529Y}.
Both theory and numerical simulations predict that a strong wind must exist in such hot accretion flows\cite{1999MNRAS.303L...1B,2012ApJ...761..130Y,2012MNRAS.426.3241N,2015ApJ...804..101Y}. The wind is of special interest not only because it is an indispensable ingredient of accretion, but perhaps more importantly, it is believed to play a crucial role in the evolution of the host galaxy via the so-called kinetic mode AGN feedback\cite{2017MNRAS.465.3291W,2019ApJ...885...16Y}.
Observational evidence for this wind, however, remains scarce and indirect\cite{2013Sci...341..981W,2016Natur.533..504C,2014MNRAS.443.2154T,2020ApJ...894...61P}.
Here we report the detection of a hot outflow from the low-luminosity active galactic nucleus in M81, based on {\it Chandra} high-resolution X-ray spectroscopy. 
The outflow is evidenced by a pair of Fe XXVI Ly$\alpha$ lines redshifted and blueshifted at a bulk line-of-sight velocity of $\pm2.8\times10^3 \rm~km~s^{-1}$ and a high Fe XXVI Ly$\alpha$-to-Fe XXV K$\alpha$ line ratio implying a plasma temperature of $1.3\times10^8$ Kelvin.
This high-velocity, hot plasma cannot be produced by stellar activity or the accretion inflow onto the SMBH.
Our magnetohydrodynamical simulations show, instead, it is naturally explained by a wind from the hot accretion flow, propagating out to $\gtrsim10^6$ times the gravitational radius of the SMBH. 
The kinetic energy and momentum of this wind 
can significantly affect the evolution of the circumnuclear environment and beyond.

\end{abstract}

Lying at a distance of $\sim$3.6 Mpc and having a systemic velocity of $-34\rm~km~s^{-1}$ [ref.\cite{1994ApJ...427..628F}], the massive spiral galaxy M81 (=NGC3031) harbors one of the nearest SMBHs, commonly known as M81*, with an estimated mass $M_{\rm BH}\approx 7\times10^{7}{\rm~M_\odot}$ [ref.\cite{2003AJ....125.1226D}]. 
The low bolometric luminosity compared to the Eddington luminosity ($L_{\rm bol}/L_{\rm Edd} \sim 3\times10^{-5}$ [ref.\cite{2014MNRAS.438.2804N}]), the existence of a radio jet\cite{2000ApJ...532..895B}, and the likely absence of a classical thin accretion disk\cite{1996ApJ...462..183H,2018MNRAS.476.5698Y}, together make M81* a prototype low-luminosity active galactic nucleus (LLAGN) powered by a hot accretion flow\cite{2014ARA&A..52..529Y}. 

The X-ray spectrum of M81* is dominated by a power-law continuum,
but also exhibits a number of prominent emission lines, in particular, highly-ionized iron lines including Fe XXV K$\alpha$ and Fe XXVI Ly$\alpha$\cite{2004ApJ...607..788D,2004A&A...422...77P,2007ApJ...669..830Y}. These lines, typically produced in a hot plasma with temperatures of $10^{7-8}$ K, hold promise for probing the accretion physics of LLAGNs.  
For this reason, M81* has been the target of the High Energy Transmission Grating (HETG) spectrometer onboard {\it the Chandra Observatory} in a total exposure of 430 ks (Extended Data Figures 1 and 2).  
These deep {\it Chandra}/HETG observations obtained high-quality X-ray spectrum of M81* from within a projected radius of $R_{\rm max} \approx$ 42 pc (equivalent to $\sim1.2\times10^7~r_{\rm g}$, where $r_{\rm g} \equiv GM_{\rm BH}/c^2$ is the gravitational radius of the black hole, $G$ is the gravitational constant and $c$ is the speed of light), providing a line-of-sight velocity resolution of 300--2000 km~s$^{-1}$ over the photon energy range of 0.5--8 keV.

We perform a {\it blind search} of emission/absorption lines in the HETG spectra against a baseline continuum model (see Methods and Extended Data Figure 3). 
Determined from a joint-fit of the HETG spectra covering 0.5--8 keV and {\it NuSTAR} spectra covering 3--79 keV, this baseline model is a single power-law with a photon-index of $1.88\pm0.01$ (quoted errors are at 90\% confidence level, unless otherwise stated), subject to Galactic foreground absorption towards M81.
It is noteworthy that the spectra show no sign of a Compton hump, suggesting that reflection from a putative accretion disk is low or absent\cite{2018MNRAS.476.5698Y}. While there is hint of a slight softening at the high-energy end of the {\it NuSTAR} spectra, our test using a power-law with an exponential cutoff finds no significant effect in the baseline continuum below 8 keV.
The blind search identifies a number of significant emission lines but no absorption lines. In particular, four emission lines are found at a most probable centroid energy of 6.40, 6.69, 6.90 and 7.05 keV, respectively (Figure~1a), all having a significance of $\gtrsim$95\% and a false detection probability of $<0.1\%$. 

The four lines are further characterized by four {\it Gaussians} to simultaneously determine the line widths and fluxes (Table 1 and Figure~1b).
The 6.40 keV line is unresolved and can be identified as the K$\alpha$ line of neutral or very weakly ionized Fe (hereafter Fe I K$\alpha$), which is expected from a truncated accretion disk or a torus of cold gas irradiated by the central LLAGN\cite{2018MNRAS.476.5698Y}.
The 6.69 keV line, having an apparent line width of $45^{+43}_{-21}$ eV, is consistent with the Fe XXV K$\alpha$ triplet. 
The 6.90 keV line, the strongest amongst the four, was previously detected in a HETG spectrum of shallower exposure and suggested to be a Fe XXVI Ly$\alpha$ line redshifted from the rest-frame energy of 6.966 keV, but the cause of the implied Doppler velocity ($\sim3000\rm~km~s^{-1}$) was unclear\cite{2007ApJ...669..830Y}.
The fourth and weakest line at 7.05 keV is unambiguously detected for the first time, but its interpretation is less straightforward.
It is tempting to associate this line with Fe I K$\beta$, which has a rest-frame energy of 7.058 keV. However, the K$\beta$-to-K$\alpha$ intensity ratio of neutral or very weakly ionized Fe has a canonical value of $\sim$0.13 [ref.\cite{2003A&A...410..359P}], whereas the observed intensity ratio between the 7.05 keV and 6.40 keV lines is $0.69\pm0.39$ (determined using a bootstrap method). 
Hence the 7.05 keV cannot be solely accounted for by Fe I K$\beta$.
A more plausible explanation for the 7.05 keV line is a blueshifted Fe XXVI Ly$\alpha$ line, pairing with the 6.90 keV line in a way roughly symmetric about the rest-frame energy.
Such a line pair signifies bulk motions of a bipolar outflow or a rotating disk/ring.

We verify this scenario with a phenomenological spectral model, which consists of a pair of collisionally-ionized, optically-thin plasma ({\it apec}\cite{2001ApJ...556L..91S} in Xspec), one accounting for the redshifted component and the other for the blueshifted component. 
The two {\it apec} components are required to have the same plasma temperature and metal abundance (fixed at solar\cite{1996ApJ...462..183H}) but exactly inverse line-of-sight velocities, and their normalizations are allowed to vary. 
A power-law and two Gaussians are also included to account for the continuum and the putative Fe I K$\alpha$ and K$\beta$ lines (adopting a fixed K$\beta$-to-K$\alpha$ ratio of 0.13). All these components are subject to the Galactic foreground absorption.  
This phenomenological model results in a good fit to the HETG spectrum over 5--8 keV, and particularly to the putative Fe lines (Figure~1c and Table 2). 
The best-fit plasma temperature, $kT_{\rm h} =11^{+3}_{-2}$ keV ($k$ is the Boltzmann constant), is mainly driven by the relative strength between Fe XXVI Ly$\alpha$ and Fe XXV K$\alpha$ and signifies an exceptionally hot plasma, compared to virial temperature, $kT_{\rm vir}=GM(<R_{\rm max}){\mu}m_{\rm p}/R_{\rm max} \approx 0.1\rm~keV$ (where $\mu \approx 0.6$ is the mean molecular weight and $m_{\rm p}$ is the mass of proton). 
The best-fit absolute line-of-sight velocity, $2.8^{+0.2}_{-0.2}\times10^{3}~\rm km~s^{-1}$, is consistent with the interpretation of the 6.90/7.05 keV lines as symmetrically redshifted/blueshifted Fe XXVI lines (the same Doppler shift that should be akin to Fe XXV K$\alpha$ is less appreciable, due to the intrinsic width of the triplet).
The redshifted and blueshifted components have a flux ratio of $1.6^{+0.2}_{-0.2}$.

Additional emission lines are found in the HETG spectrum at lower energies, in particular between 1--3 keV (Extended Data Figures 4 and 5), which are mainly identified as the helium-like and hydrogen-like transitions of $\alpha$ elements (Si, S and Ar).  
The pair of the 11-keV {\it apec} components, however, cannot simultaneously account for these low-energy lines, since most low-Z elements would become fully ionized at this high temperature. 
Moreover, despite a higher velocity resolution at these low-energy lines, none of them shows a line broadening or Doppler shift comparable to that of Fe XXVI Ly$\alpha$,
indicating that the line-of-sight velocity of $\pm$2800 km~s$^{-1}$ is only related to the 11-keV plasma. 
Hence we introduce a third {\it apec} component to account for the He-like and H-like lines of low-Z elements.
It is also necessary to introduce two additional Gaussian components to account for the neutral K$\alpha$ line of Si and Ar.
This multi-component model leads to a reasonably good fit to the HETG spectrum over the 0.5--8 keV range (Table 2), in which
the third {\it apec} component has a temperature of $kT_{\rm l}=0.9^{+0.1}_{-0.1}$ keV 
and an unabsorbed 0.5-10 keV luminosity of 
$3.3(\pm0.5)\times10^{38}\rm~erg~s^{-1}$. 
This 0.9-keV plasma is unseen in the HETG zeroth-order spectrum of an annular region immediate outside $R_{\rm max}$ (see Methods and Extended Data Figure 6), suggesting that it is spatially confined and may trace a circumnuclear diffuse hot gas.

The high-velocity Fe lines can in principle be produced in stellar activities such as young supernova remnants (SNRs) or massive star binaries with strong colliding winds. However, the X-ray luminosity of the 11-keV plasma,
$3.8\times10^{39}\rm~erg~s^{-1}$, is exceedingly high for a SNR; only SNe younger than $\sim$10$^4$ days can have X-ray luminosities $\gtrsim10^{39}\rm~erg~s^{-1}$ [ref.\cite{2012MNRAS.419.1515D}], but such a recent and nearby SN could hardly have been missed by astronomers\cite{1993Natur.364..600S},
consistent with our estimated SN birth rate of $\lesssim 5\times10^{-5}\rm~yr^{-1}$ within $R_{\rm max}$. 
Similarly, the observed X-ray luminosity and plasma temperature are too high for a colliding wind binary\cite{2012ASPC..465..301G}. 
Unresolved X-ray binaries are also insufficient to account for the observed X-ray flux, given the moderate amount of stellar mass and star formation rate within $R_{\rm max}$ (Methods).
Therefore, we conclude that stellar activities cannot be responsible for the high-velocity 11-keV plasma.  

Photoionization by the LLAGN can be safely ruled out as the cause of the highly-ionized Fe lines, because the required irradiating luminosity is orders of magnitude higher than the observed X-ray luminosity of M81* (Methods and Extended Data Figure 7). The high-velocity 11-keV plasma is also highly unlikely to be a jet-driven outflow, because a jet can only drive a fast outflow by a shock ahead of the ``jet head'', 
leaving little effect on the kinetics of the gas behind\cite{2018MNRAS.473.1332G}. 
The steady jet of M81* should have long passed the region ($\lesssim$ 42 pc) probed by the HETG observations.

Another possibility is that the highly-ionized Fe lines may be produced in the accretion inflow, which, for an LLAGN like M81*, consists of a truncated thin disk plus an inner hot accretion flow\cite{2014ARA&A..52..529Y} (see illustration in Figure 2a). 
In the thin disk, the rotational velocity $v_\phi \sim(GM_{\rm BH}/r)^{\frac{1}{2}}$ can reach $\gtrsim3000\rm~km~s^{-1}$ at radii $r\lesssim10^4~r_{\rm g}$. While this is compatible with the estimated truncated radius\cite{2018MNRAS.476.5698Y}, $r_{\rm tr}$ $\sim$ a few $10^3~r_{\rm g}$, the disk temperature can only reach $\sim10^6\rm~K$, which is too low to be consistent with the observed Fe XXVI-to-XXV line ratio. 
Inside $r_{\rm tr}$, the thin disk is replaced by a geometrically-thick hot inflow, where plasma temperatures become much higher ($\sim10^9 [2\times10^3~r_{\rm g}/r]\rm~K$, [ref.\cite{2014ARA&A..52..529Y}]). 
In this case, our numerical simulation of the hot accretion flow in M81* predicts an equivalent width of Fe XXVI L$\alpha$ much lower than the observed value ($\sim$110 eV), because the temperature of the accretion flow is so high that the plasma is too strongly ionized (see Methods and Extended Data Figure 8). 
A larger $r_{\rm tr}$  would in principle lead to stronger Fe XXVI L$\alpha$; however, the correspondingly decreased rotational velocity would result in a reduced Doppler shift inconsistent with the observed double-peak line profile. 
Therefore the possibility that the high-velocity Fe lines  originate in the accretion inflow,  either the truncated thin disk or the inner hot accretion flow, can be ruled out.

This leaves a bipolar wind launched from the hot accretion flow as the most plausible origin of the high-velocity Fe lines.
Such a wind, schematically illustrated in Figure 2a, is a robust prediction by theory\cite{1999MNRAS.303L...1B} and numerical simulations of hot accretion flows\cite{2012ApJ...761..130Y,2012MNRAS.426.3241N,2015ApJ...804..101Y}.  
Different from a relativistic jet, the wind originates from a large range of radius in the accretion flow, having a much wider opening angle and much smaller radial velocity.
The temperature of the wind decreases from that of the hot accretion flow due to adiabatic expansion when propagating outward, and thus may be suited for producing the observed Fe lines.

We have performed 2.5-dimensional magnetohydrodynamical (MHD) simulations of the wind launched from the hot accretion flow in M81* with $r_{\rm tr}=3000~r_{\rm g}$ to obtain the spatial distributions of density, temperature, velocity and magnetic field (Figure 2b-d), which in turn allow us to generate synthetic X-ray spectrum to compare with the HETG spectrum (see Methods).
We find that self-absorption in the Fe XXVI Ly$\alpha$ and XXV K$\alpha$ lines is negligible for the wind, which has an equivalent column density of $\lesssim10^{22}\rm~cm^{-2}$.
A free parameter in this synthetic spectrum is the inclination angle between the jet axis and the light-of-sight. 
Figure~1d illustrates a synthetic spectrum with an inclination angle of $15^\circ$, a value consistent with that inferred from the plane of motion of circumnuclear warm gas\cite{2003AJ....125.1226D}
and also compatible with the upper limit of $56^\circ$ inferred from radio observations of moving jet knots\cite{2016NatPh..12..772K}.
The synthetic spectrum exhibits a double-peak Fe XXVI L$\alpha$ profile and a high XXVI-to-XXV flux ratio, both in reasonable agreement the observed spectrum.
The redshifted and blueshifted components in the synthetic spectrum have nearly equal amplitudes, which is a consequence of the mirror symmetry about the equatorial plane of the accretion flow assumed in our simulation. In reality, the wind is usually intrinsically asymmetric above and below the equatorial plane, easily causing mild (at a level of $\sim$25\%) inequality in the gas density and hence the $\sim$60\% difference between the observed redshifted/blueshifted components on timescales significantly shorter than the wind dynamical time of $10^4~r_{\rm g}/v_r\sim 10$ yr, with $10^4~r_{\rm g}$ being the region where most of the Fe XXVI line emission originates and $v_r$ the wind radial velocity. It is noteworthy that the moderate inequality between the blueshifted and redshifted Fe lines does not affect our arguments and conclusions against the aforementioned alternative origins of a high-velocity hot plasma.

Theoretically\cite{2015ApJ...804..101Y}, the outflow rate of the hot wind should be nearly equal to the inflow rate of the hot accretion flow (also of the truncated thin disk) at $r_{\rm tr}$. 
The outflow rate can be estimated using the density and radial velocity information from the wind simulation, which is found to be $\sim 2\times10^{-3}\rm\ M_{\odot}\ yr^{-1}$. 
Indeed, this value is in good agreement with  
the estimated mass inflow rate ($4\times10^{-3}\rm\ M_{\odot}\ yr^{-1}$) of the warm ionized gas at a radius of $20\rm~pc$ ($\sim6\times10^6~r_{\rm g}$)\cite{2011MNRAS.413..149S}, 
%Given some mass loss of the gas due to the weak wind from the truncated thin disk, 
lending strong support to the wind model. 
The associated wind kinetic power is $\sim 2\times10^{40}\rm\ erg\ s^{-1}$, equivalent to 10\% of the bolometric luminosity of M81*\cite{2014MNRAS.438.2804N}. 
On the other hand, the wind momentum flux, $\sim6\times10^{31}\rm~g~cm~s^{-2}$, is three times of the radiation flux ($L_{\rm bol}/c \sim2\times10^{31}\rm~g~cm~s^{-2}$).

The bulk of the wind momentum and kinetic energy should be deposited into the interstellar medium (ISM) due to the wide opening angle of the wind, thus providing an effective feedback to the host galaxy and regulating the growth of the SMBH\cite{2017MNRAS.465.3291W,2019ApJ...885...16Y}. 
This is supported by the existence of the 0.9-keV component in the X-ray spectrum, which is likely the result of the outward propagating wind shock-heating the ISM. Numerical simulations of wind-ISM interaction in a realistic galactic environment\cite{2019ApJ...885...16Y} do find that winds with a similar power from a hot accretion flow stop roughly at a few times $10^6~r_{\rm g}$, which is consistent with the extent of the 0.9-keV gas. The internal energy of this 0.9-keV gas, %$2{\pi}r_{\rm out}^3\bar{n}kT_{\rm l} \lesssim1.2\times10^{53}\rm~erg$, 
estimated to be $\lesssim1.2\times10^{53}\rm~erg$,
can be converted from the wind kinetic energy in $2\times10^4$ yr, comparable to the dynamical time of the wind. 
Such a wind-ISM interaction is also evidenced by a previously identified blob of warm ionized gas located at $\sim$ 10 pc south of M81*\cite{2015A&A...576A..58R}.

\clearpage

\begin{figure*}
\hspace{-1.5cm}
\centering\includegraphics[width=1\linewidth]{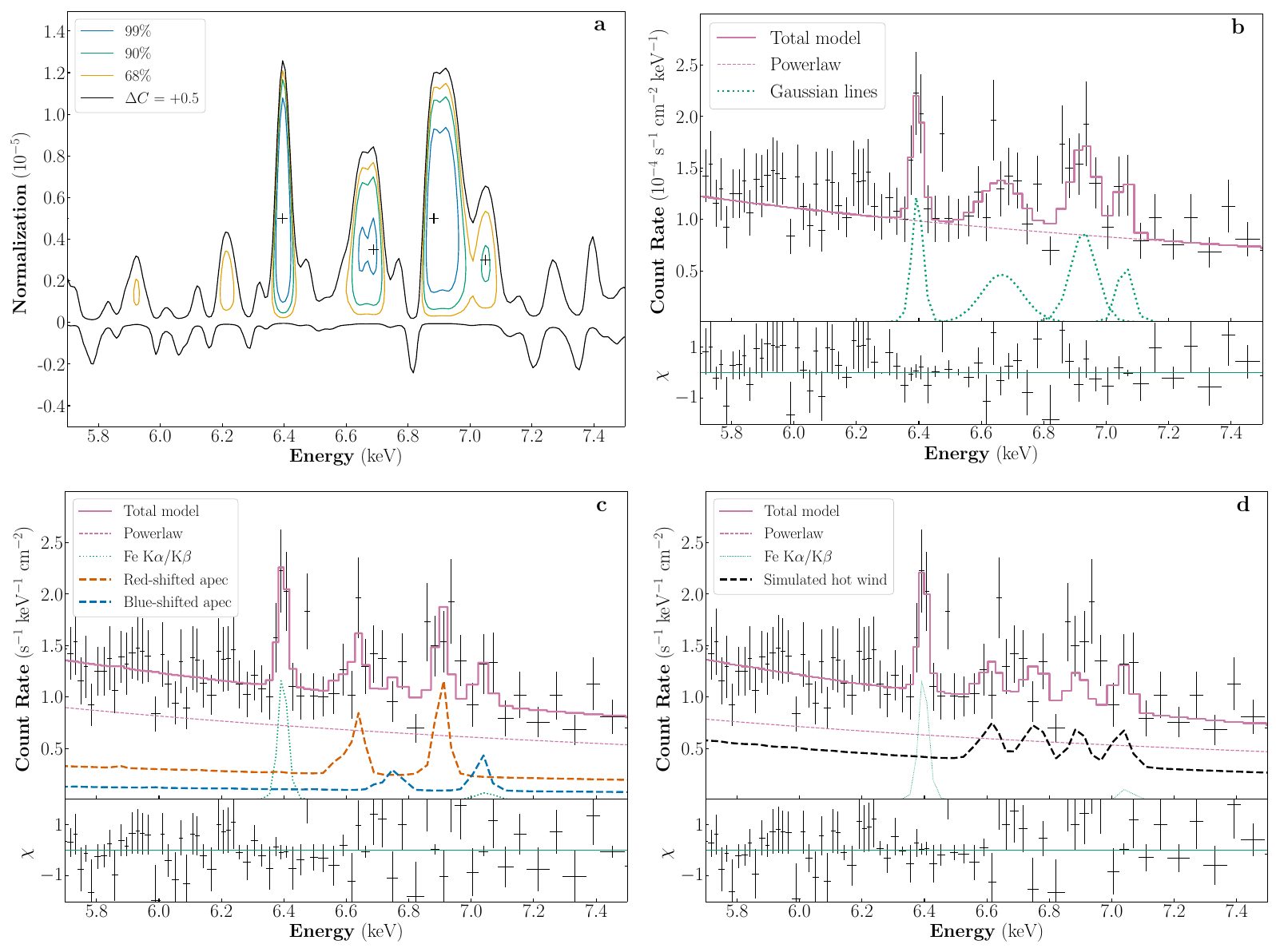}
\caption{{\bf Detection and characterization of highly-ionized Fe lines.} {\bf a,} Blind line search of the High Energy Grating (HEG) 1st-order spectrum identifies four significant Fe lines. Blue, green and orange contours indicate the confidence level 99\%, 90\% and 68\%,  respectively of a test line, according to the differential $C$-stat value $\Delta C$ against the baseline continuum model. Black contour denotes where $\Delta C=+0.5$. The '+' sign marks the most probable centroid energy and normalization for each line. 
{\bf b-d,} The HEG 1st-order spectrum (black crosses), binned to achieve a signal-to-noise-ratio greater than 3 for illustration. Error bars are of 1\,$\sigma$. The purple solid line indicates the total model. The ratio of residual/error $\chi$ is shown in the bottom of each panel.
In {\bf b,} Each of the four Fe lines is fitted by a Gaussian profile (green dotted line), on top of the baseline power-law model (purple dotted line).
In {\bf c,} The model consists of a power-law (purple dotted lines), two Gaussians accounting for the neutral Fe K$\alpha$ (6.40 keV) and K$\beta$ (7.058 keV) lines (green dotted line), and two {\it apec} components with a line-of-sight velocity of $\pm2800\rm~km~s^{-1}$ (blue and orange dashed lines).
In {\bf d,} The two {\it apec} components are replaced by a synthetic wind spectrum (black dashed line) with a viewing angle of $15^{\circ}$.}
\end{figure*}
\clearpage

\begin{figure*}
\centering\includegraphics{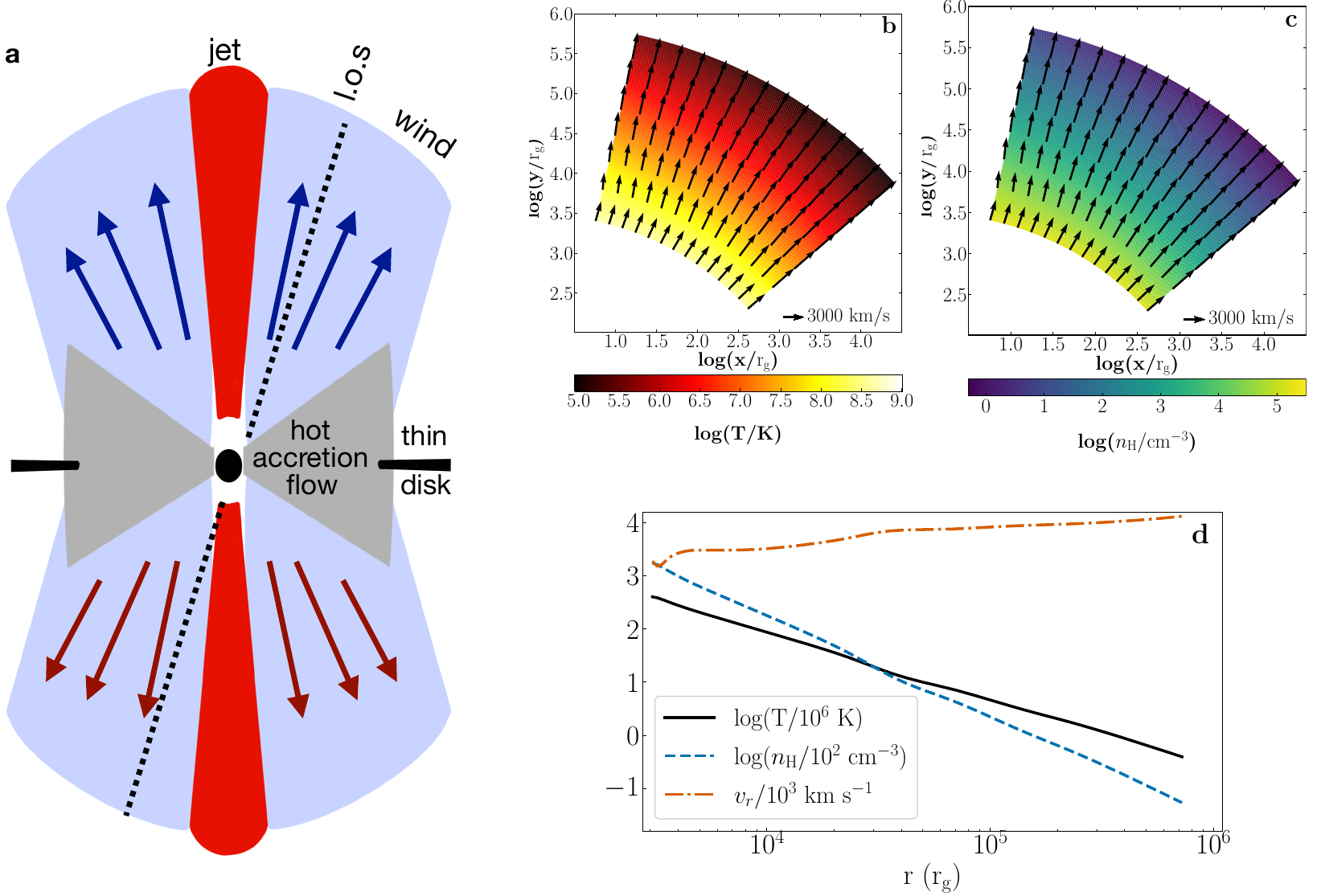}
\caption{{\bf The hot wind model.} {\bf a,} 
Schematic diagram of the accretion inflow and wind around the SMBH (represented by the black circle). A cold, thin disk occupies the outer horizontal region (shown in black) and is truncated at a radius $r_{\rm tr} \sim 3\times10^3~r_{\rm g}$, within which the accretion flow becomes hot and geometrically-thick (grey). 
A collimated, relativistic jet is present along the polar region (red) and may contribute to or even dominate the X-ray continuum. 
A hot wind is launched from the hot accretion flow and fills the region (light blue) between the jet and the accretion flow, where
optically-thin thermal X-ray emission is produced. 
Emission lines generated in the hot wind may be seen redshifted and blueshifted depending on the viewing angle (l.o.s).
The various physical components are not to scale. 
{\bf b,} Plasma temperature, and {\bf c,} density distributions from the MHD simulation of the hot wind, overlaid by vectors representing the radial velocity. 
The y-axis marks the direction of jet and the x-axis marks the direction of the equatorial plane. The MHD simulation covers a range from $3\times10^3\rm~r_g$ to $10^6\rm~r_g$ in radius and from 5$^\circ$ to 50$^\circ$ with respect to the jet.
The density has been normalized to match the observed X-ray spectrum.
{\bf d,} The radial distribution of emissivity-weighted temperature (solid black), number density (dashed blue) and radial velocity (dash-dotted vermillion) of the wind.}
\end{figure*}

\clearpage
\begin{table}
\centering
\caption{{\bf Detection of the Fe lines}}
\label{tab:line}
\medskip
\begin{threeparttable}
\begin{tabular}{cccccc}
\hline
$E_0$ & Significance & $E_{\rm c}$ & $\sigma_{\rm E}$&  $F$ & EW \\
$\rm keV$ & & $\rm keV$ & $\rm eV$ & $\rm \ erg~s^{-1}~cm^{-2}$ & $\rm eV$\\
(1) & (2) & (3) & (4) & (5) & (6)\\
\hline
$6.395$ & \textgreater\ $99.99\%$ & $6.397^{+0.007}_{-0.008}$ & \textless\ $57$ & $5^{+2}_{-2}\times10^{-14}$ & $43.4^{+0.4}_{-0.4}$ \\
%\hline
$6.688$ & $99.86\%$ & $6.666^{+0.033}_{-0.032}$ & $45^{+43}_{-21}$ & $7^{+3}_{-3}\times10^{-14}$ & $57.4^{+0.9}_{-0.7}$\\
%\hline
$6.902$ & \textgreater\ $99.99\%$ & $6.914^{+0.022}_{-0.021}$ & $38^{+22}_{-14}$ & $9^{+4}_{-3}\times10^{-14}$ & $80.8^{+0.4}_{-1.2}$ \\
%\hline
$7.050$ & $94.76\%$ & $7.046^{+0.060}_{-0.065}$ & \textless\ $120$ & $4^{+3}_{-3}\times10^{-14}$ & $31.2^{+0.6}_{-0.7}$\\
\hline
\end{tabular}
\begin{tablenotes}
\footnotesize
\item Notes -- (1) Most probable line energy determined from blind line search; 
(2) Significance level of the line; 
(3)-(6) Line centroid, width, flux and equivalent width determined from a Gaussian model. Error bars are at 90\% confidence level; upper limits in the line width are given at 3\,$\sigma$ level.
\end{tablenotes}
\end{threeparttable}
\end{table}

\clearpage
\begin{table}
\centering
\caption{{\bf Broad-band Spectral Fit Results}}
\begin{threeparttable}
\scriptsize
\begin{tabular}{c|cccccccccc}
\hline
Data \& Model & $C$/d.o.f & $\Gamma$ & $N_{\rm PL}$ & $kT_{\rm h}$ & $V_{\rm w}$ & $N_{\rm h1}$ & $N_{\rm h2}$ & $kT_{\rm l}$ & $N_{\rm l}$ & $N_{\rm wind}$\\
(1) & (2) & (3) & (4) & (5) & (6) & (7) & (8) & (9) & (10) & (11) \\
\hline
$PL$ & &  & & & & & & & &\\
\hline
HEG (1-6/7.5-8 keV)& 4388/4142 & \multirow{4}*{$1.88^{+0.01}_{-0.01}$} & $3.42^{+0.03}_{-0.03}$ & - & - & - & - & - & - & - \\ %\multirow{4}*{-} & \multirow{4}*{-} & \multirow{4}*{-} & \multirow{4}*{-} & \multirow{4}*{-} & \multirow{4}*{-}\\
MEG (0.5-5 keV)& 4586/4557 & ~ & $3.55^{+0.02}_{-0.02}$ & - & - & - & - & - & - & - \\
FPMA (3-6/7.5-79 keV)& 3770/3452 & ~ & $6.73^{+0.09}_{-0.09}$ & - & - & - & - & - & - & - \\
FPMB (3-6/7.5-79 keV)& 3546/3456 & ~ & $6.87^{+0.09}_{-0.09}$ & - & - & - & - & - & - & - \\
\hline
$PL+apec_{\rm h}\times2$ &  & & & & & & & & & \\
\hline
HEG (5-8 keV)& 408/363 & 1.88 & $3.02^{+0.11}_{-0.10}$ & $11^{+3}_{-2}$ & $2.8^{+0.2}_{-0.2}$ & $7^{+4}_{-3}$ & $4^{+4}_{-2}$ & - & - & - \\
\hline
$PL+wind$ &  & & & & & & & & &\\
\hline
HEG (5-8 keV)& 418/365 & 1.88 & $2.07$  & - & - & - & - & - & - & $0.6$ \\
\hline
$PL+apec_{\rm h}\times2+apec_{\rm l}$ & & & & & & & & & & \\
\hline
HEG (1-8 keV)& 4541/4302 & 1.88 & \multirow{2}*{$2.94^{+0.14}_{-0.09}$} & \multirow{2}*{$16^{+7}_{-3}$} & \multirow{2}*{$2.9^{+0.2}_{-0.1}$} & \multirow{2}*{$10^{+5}_{-4}$} & \multirow{2}*{$5^{+4}_{-3}$} & \multirow{2}*{$0.9^{+0.1}_{-0.1}$} & \multirow{2}*{$1.1^{+0.2}_{-0.2}$} & - \\
MEG (0.5-5 keV)& 4498/4551 & 1.88 & ~ & ~ & ~ & ~ & ~ & ~ & ~ & -\\
\hline
\end{tabular}
\small 
Notes --  (1) Fitted spectral set (with the energy band in parenthesis) and the adopted model; (2) Value of the Cash statistic and degree of freedom; (3) Photon-index of the power-law; 
(4) Normalization of the power-law component in units of $10^{-3}\rm~photon~s^{-1}~cm^{-2}~keV^{-1}$ at 1 keV;
(5) Plasma temperature in units of keV, same for the two $apec_{\rm h}$ components; (6) Absolute line-of-sight velocity of the two $apec_{\rm h}$ components, in units of $10^3\rm~km~s^{-1}$; 
(7)-(8) Normalization of the redshifted  (h1) and blueshifted (h2) $apec_{\rm h}$ components, in units of $10^{-4}\rm~cm^{-5}$;
(9) Plasma temperature of $apec_{\rm l}$, in units of keV;
(10) Normalization of $apec_{\rm l}$, in units of $10^{-4}\rm~cm^{-5}$;
(11) Normalization of the hot wind model with an viewing angle of $15^{\circ}$, in units of $10^{-4}\rm~cm^{-5}$.
Quoted errors are at 90\% confidence level.
\label{tab:my_label}
\end{threeparttable}
\end{table}

\clearpage

\begin{methods}

{\bf X-ray data}\\
In this study we utilize two sets of X-ray data: i) {\it Chandra} ACIS/HETG 1st-order grating spectra to resolve emission lines at a velocity resolution of 300--2000 km~s$^{-1}$ over energy range of 0.5--8 keV, and ii) {\it NuSTAR} spectra to constrain the continuum over 3--79 keV.
A log of the X-ray data is given in Extended Data Figure~1.

{\it Chandra} observed M81* with the combined operation of its Advanced CCD Imaging Spectrometer (ACIS) and High Energy Transmission Grating (HETG) in 15 epochs between February 24, 2005 and August 12, 2006. 
The publicly available data were reprocessed using CIAO v4.9 and calibration files (CALDB v4.8.2) and following the standard pipeline. 
Time intervals of high particle background were filtered, resulting in a total cleaned exposure of 429.2 ks. 
The $\pm$1st-order spectra of both the High Energy Grating (HEG) and Medium Energy Grating (MEG) were extracted for each observation, using the CIAO tools {\sl tg\_resolve\_events} and {\sl tgextract}.
For the source spectra, we adopted the default cross-dispersion half-width of 2.4$''$ (equivalent to a projected radius $R_{\rm max} \approx$ 42 pc), which ensures the same enclosed energy fraction ($\sim$97\%) for different wavelengths, according to the Proposer's Observatory Guide (https://cxc.cfa.harvard.edu/proposer/POG/). 
We have also tested a smaller cross-dispersion half-width of 1.5$''$ and found that the main spectral properties, in particular the Fe lines, are unaffected.
The background spectra were extracted from a default adjacent region of 19.1$''$ full-width below and above the source region. 
As shown in Extended Data Figure~2, the 1st-order arms have a varied position angle among the 15 observations, and in some cases intercept with off-nucleus point sources (mostly X-ray binaries belonging to M81\cite{2011ApJ...735...26S}).
Hence we carefully examined each observation to mask out any contaminating sources. 
We then coadded the spectra of $\pm$1st-order and of individual observations to form a combined HEG/MEG spectrum, along with the exposure-weighted ancillary response files (ARFs) and redistribution matrix files (RMFs). 
The corresponding background spectra were similarly coadded. 
Thanks to the relatively high flux of M81*, the background contributes less than 2\% to the source spectra.

{\it NuSTAR} observed M81* between May 18--20, 2015, covering a duration of 342 ks. 
Data reduction was performed using NuSTARDAS v1.7.1 and the HEASOFT/FTOOLS v6.21.1. 
After applying {\it nupipeline} to obtain cleaned event files with a net exposure of 209.0 ks, we extracted the spectrum of M81* over 3--79 keV from both Focal Plane Modules A and B (FPMA \& FPMB). 
The source region was defined as a circle with a radius of 100$''$, approximately enclosing 90\% of the counts from a point source, while 
the background was extracted from an annulus with an inner radius of 105$''$ and an outer radius of 175$''$ (Extended Data Figure 2). 
We note that the {\it NuSTAR} spectral extraction region encloses 61 off-nucleus X-ray sources as detected by {\it Chandra}\cite{2011ApJ...735...26S}. Their collective 0.5--10 keV flux is less than 2\% of the flux of M81* in the same energy range.
Therefore, it is expected that the contribution of these off-nucleus sources is negligible in the {\it NuSTAR} spectra. 

\noindent{\bf Baseline continuum model}\\
Spectral fitting is carried out with XSPEC v.12.9.1, %(https://heasarc.gsfc.nasa.gov/docs/xanadu/xspec/index.html), 
which employs ATOMDB v.3.0.9 (http://www.atomdb.org/) for the modelling of atomic lines. 
To preserve the maximally possible spectral resolution, we group the coadded spectra to have at least one count per bin and employ the Cash statistic ($C$-stat)\cite{1979ApJ...228..939C} in the fit. 

Previous studies have shown that the X-ray spectrum of M81* is dominated by a single power-law\cite{2004ApJ...607..788D,2004A&A...422...77P,2007ApJ...669..830Y,2018MNRAS.476.5698Y}, likely arising from a synchrotron jet and/or Comptonization in the hot accretion flow. 
Hence an absorbed power-law ({\it tbabs*powerlaw} in XSPEC) is adopted as the baseline continuum model. The {\it Chandra}/HETG spectra over 0.5--8 keV and the {\it NuSTAR} spectra over 3--79 keV are jointly fitted to provide the tightest constraint on the photon-index (Table 2 and Extended Data Figure~3). The normalization is allowed to vary, accounting for the different enclosed energy fractions and possible flux variability between the {\it Chandra} and {\it NuSTAR} spectra. 
The absorption column density ($N_{\rm H}$) is tied among the spectra and required to have a minimum value of $6.4\times 10^{20}~\rm cm^{-2}$ to be consistent with the Galactic foreground absorption\cite{2005A&A...440..775K}. 
In this step we exclude the energy range of 6--7.5 keV that cover the Fe lines known to present in the X-ray spectrum\cite{2004ApJ...607..788D,2004A&A...422...77P,2007ApJ...669..830Y}. 
The baseline model gives a best-fit photon-index $\Gamma = 1.879\pm0.005$ with $C$-stat = 12117.5 for 11745 degrees of freedom. 

It is noteworthy that significant X-ray variability is seen in M81* on timescales from hours\cite{2018MNRAS.476.5698Y} to years\cite{2004ApJ...601..831L}. We have examined the spectra of the 15 individual HETG observations spanning 18 months and found that the best-fit photon-index is consistent with each other within 95\% confidence level, in agreement with previous work\cite{2010ApJ...720.1033M}.
Similarly, we find that the photon-index of M81* remained constant through the 209-ks exposure of the {\it NuSTAR} observation, despite the fact that its 3--79 keV flux increased by 30\% in the last 40 ks [ref.\cite{2018MNRAS.476.5698Y}]. 

The {\it NuSTAR} spectra suggest a slight softening in the high-energy tail. Thus we also consider an absorbed power-law with an exponential cutoff ({\it tbabs*cutoffpl} in XSPEC) as the continuum model. 
This results in an improvement of $C$-stat, $\Delta C=29$, for one more degree of freedom compared to the absorbed power-law model.
The best-fit cutoff energy, $E_{\rm cut}=242^{+109}_{-58}\rm~keV$, is  consistent with previous work\cite{2018MNRAS.476.5698Y}. 
In view of the presence of the putative Fe I K$\alpha$ and K$\beta$ lines, yet an alternative treatment of the continuum can be obtained by using a disk reflection model ({\it pexmon} in Xspec), which self-consistently accounts for fluorescent neutral Fe lines in response to a central power-law source. We have fitted such a model to the joint HEG and NuSTAR spectra. The inclusion of the latter warrants a strong constraint on the reflection fraction. 
The inclination angle of the irradiated disk is fixed at 45$^\circ$, but the exact value of this parameter does not significantly affect the result. 
The fit finds a photon-index of $1.83\pm0.01$, a cut-off energy of $85\pm5$ keV, and a low reflection fraction with a 3-$\sigma$ upper limit of 0.27, which are broadly consistent with ref.\cite{2018MNRAS.476.5698Y}.
We find that neither the inferred high-energy cutoff nor the weak reflection has a significant effect on the Fe lines, hence we keep the absorbed power-law as our standard baseline model for the following blind line search.

\noindent{\bf Blind line searching and likelihood test}\\
To identify emission/absorption lines in the HEG and MEG spectra, we perform a {\it blind search} following the method described in ref.\cite{2010A&A...521A..57T}. 
Specifically, the putative line is modeled by a {\it Gaussian} profile with a line width equaling half of the natal spectral bin (0.0056 \AA\ for HEG and 0.0111 \AA\ for MEG).
We then add the Gaussian line to the baseline continuum model (i.e., the best-fit absorbed power-law) and progressively shift the line centroid ($E_0$) through the coadded HEG/MEG spectra, at each step allowing the normalization ($N_0$) of the Gaussian line to vary. 
For each pair of $E_0$ and $N_0$, $C$-stat is evaluated and differentiated with that of the baseline model (i.e., without the line) to obtain the value of ${\Delta}C$, which indicates the significance of fit improvement.
A contour plot of ${\Delta}C$ can thus be generated over the $N_0 - E_0$ parameter space, as shown in Figure~1 and Extended Data Figure 4. 
The confidence level of the line can be calculated according to ${\Delta}C$, given the fact that $C$-stat is asymptotically distributed as $\chi^{2}$.
A number of emission lines are thus identified to have a significance $\geq 90\%$ in the coadded MEG or HEG spectrum, or both (Table 1 and Extended Data Figure 5), but no significant absorption line is found.
We further require that any line found between 1--3 keV, where the MEG and HEG spectra have comparable line sensitivity, is considered {\it real} only if its flux or 2\,$\sigma$ upper limit were consistent between the MEG and HEG spectra. 
Among these lines, three are new detections (the forbidden and resonant transitions of S XV K$\alpha$, and S XVI Ly$\alpha$), while the remaining six lines have been previously reported\cite{2007ApJ...669..830Y}.

We test the likelihood that the putative emission line arises from pure statistical fluctuations using a bootstrap method\cite{2010A&A...521A..57T}.
%We focus on the four Fe lines at 6.40, 6.69, 6.90 and 7.05 keV, respectively.
First, a fake continuum-only spectrum is generated according to the baseline model. This is achieved by using the {\it fakeit} command in Xspec and taking into account the instrumental response (ARFs and RMFs) and the exact exposure of the spectra. 
The fake spectrum is then used to calculate the significance of the test line in a way identical to that applied for the real spectrum.  
This procedure is repeated 1000 times, each time the significance level is recorded.
The false detection probability of the line in question is then counted as the fraction of fake spectra in which the value of ${\Delta}C$ is equal to or larger than that of the real spectrum. It is found that the false detection rate is less than 0.1\% for all emission lines identified. 

\noindent{\bf Zeroth-order spectrum of the circumnuclear region}\\
The HETG zeroth-order image, i.e., X-ray photons directly captured by the detector without dispersion by the grating system, can provide useful constraint on the spatial extent of the thermal components found in the 1st-order dispersed spectra. 
The strong flux of M81* caused significant pile-up (i.e., more than one photons are registered within one readout frame of 3.2 sec) in the central few pixels. Therefore, we extracted a spectrum from an annulus around M81* with an outer radius of 5$''$\ and an inner radius of 2.5$''$, which is complement to the 1st-order spectra. The corresponding background spectrum was extracted from a concentric annulus between 5$''$\ and 7$''$. Discrete sources detected within the source and background regions were excluded. 
The spectra from the 15 individual observations were then coadded. 
The background-subtracted, coadded spectrum can be well fitted with a power-law model subject to Galactic foreground absorption (Extended Data Figure~6). 
The best-fit photon-index, $\Gamma=1.7\pm0.1$, along with the unabsorbed 0.5--10 keV luminosity of $3.8\times10^{38}\rm~erg~s^{-1}$, is consistent with the collective emission from unresolved X-ray binaries plus the PSF-scattered photons from M81*.
While an additional thermal component is not formally required by the spectrum, we added an {\it apec} model to the fit, fixing the plasma temperature at $0.9$ keV and the abundance at solar. 
The normalization was adjusted to obtain the 3\,$\sigma$ upper limit allowed by the spectrum. This corresponds to an unabsorbed 0.5--10 keV luminosity of $2.7\times10^{37}\rm~erg~s^{-1}$, which is about an order of magnitude lower than that of the 0.9-keV component found in the 1st-order spectrum (and at least a factor of 40 lower in surface brightness), suggesting that the latter is spatially confined and traces a circumnuclear hot gas.

\noindent{\bf Black hole mass, enclosed stellar mass and star formation rate}\\
The mass of the SMBH in M81 has been estimated in several studies. 
Based on the kinematics of circumnuclear ($\lesssim$10 pc) ionized gas measured by HST/STIS spectroscopy, Ref.\cite{2003AJ....125.1226D} obtained $M_{\rm BH} = 7^{+2}_{-1}\times10^{7}\rm~M_\odot$ along with an inclination angle of $i \approx 14^{\circ}$ for the plane of gas motion.
A value of $M_{\rm BH} =6(\pm20\%)\times10^7\rm~M_\odot$ was reported by Ref.\cite{2000AAS...197.9203B} based on stellar kinematics. 
Ref.\cite{2011MNRAS.413..149S} obtained $M_{\rm BH}=5.8^{+3.8}_{-2.1}\times10^{7}\rm~M_\odot$ (rescaled to our adopted distance of 3.6 Mpc) based on the black hole mass--stellar velocity dispersion ($M_{\rm BH} - \sigma_{*}$) relation\cite{2013ARA&A..51..511K}. 
We adopt $M_{\rm BH} =7\times10^{7}\rm~M_\odot$ as our fiducial black hole mass. 

The stellar mass projected within $R_{\rm max}$ is estimated to be $M_* \approx 3\times10^{8}\rm~M_\odot$, based on a S{\'e}rsic profile fitted to the surface brightness distribution of the stellar bulge\cite{2003AJ....125.1226D}. 
The birth rate of Type Ia supernovae associated with this stellar mass is $R^{\rm Ia}_{\rm SN} \approx 2.0\times10^{-5}\rm~yr^{-1}$, according to the empirical relation of ref.\cite{2005A&A...433..807M}.
Based on the measured circumnuclear H$\alpha$ luminosity\cite{1996ApJ...462..183H}, $L_{\rm H\alpha} \approx 3\times10^{38}\rm~erg~s^{-1}$, we can estimate an upper limit in the star formation rate, SFR$\lesssim 0.002\rm~M_\odot~yr^{-1}$, according to the empirical ${\rm SFR} -L_{\rm H\alpha}$ relation\cite{1998ARA&A..36..189K}. This can be translated to a birth rate of core-collapse supernovae, $R^{\rm cc}_{\rm SN} \lesssim 2.5\times10^{-5}\rm~yr^{-1}$, for a Chabrier/Kroupa initial mass function\cite{2016MNRAS.456.2537R}.

The above estimates on stellar mass and SFR in turn provide an upper limit in the X-ray luminosity contributed by both the old and young stellar populations within $R_{\rm max}$, $L_{\rm 2-10~keV}^{\rm star} \lesssim 3\times10^{37}\rm~erg~s^{-1}$, according to the empirical relation\cite{2010ApJ...724..559L} of $L_{\rm 2-10~keV}^{\rm star} = \alpha M_* + \beta {\rm SFR}$, where $\alpha = (9.05\pm0.37)\times10^{28}{\rm~erg~s^{-1}~M_\odot}^{-1}$ and $\beta = (1.62\pm0.22)\times10^{39}{\rm~erg~s^{-1}~(M_\odot~yr^{-1})}^{-1}$. 
Clearly, the expected stellar contribution is too low to account for  the observed 2--10 keV luminosity of M81* ($1.7\times10^{40}{\rm~erg~s^{-1}}$), and in fact falls short for just the 11-keV {\it apec} component ($3\times10^{39}{\rm~erg~s^{-1}}$). 
Star formation activity also cannot account for the 0.9-keV thermal component, which has a 0.5--2 keV luminosity of $\sim3\times10^{38}{\rm~erg~s^{-1}}$, significantly higher than the predicted value of $L_{\rm 0.5-2~keV}^{\rm SFR} = 9\times10^{36}({\rm SFR}/0.002\rm~M_\odot~yr^{-1})erg~s^{-1}$ based on the empirical relation of ref.  \cite{2003A&A...399...39R}.

\noindent{\bf Ruling out AGN photoionization}\\
We consider the possibility that 
the highly-ionized Fe lines are originated in a photoionized gas irradiated by the central black hole. 
Using the publicly available photoionization code, {\it Cloudy} (version c17.02)\cite{2017RMxAA..53..385F},
we first calculate the ionization state of various elements in an isotropic and uniform gas cloud photoionized by a central point source, which has an intrinsic X-ray spectrum identical to the best-fit power-law model of M81*, i.e., with a photon-index of 1.88 and 0.5--10 keV luminosity of $5\times10^{40}\rm~erg~s^{-1}$. 
The inner irradiated surface of the cloud is fixed at $r_{\rm in}=0.001\rm~pc$ ($\sim300~r_{\rm g}$) from the central source. 
The ionization state of the cloud is then determined by the dimensionless ionization parameter, $U_{\rm X}=\int_{2}^{10}\frac{L_{\nu}}{h\nu}{\rm d}{\nu}/(4\pi r_{\rm in}^2cn_{\rm c})$, and the column density $N_{\rm c} = n_{\rm c}(r_{\rm out}-r_{\rm in})$, where $n_{\rm c}$ is the constant cloud density (We have also tested a radially-decreasing density distribution and found that our conclusion below remains unchanged.). 

We have explored the plausible parameter space of $N_{\rm c}$ from $10^{20}$ to $10^{23}\rm~cm^{-2}$ and $log(U_{\rm X})$ from -0.5 to 5.0.
The outward luminosity of a given atomic line, in particular Fe XXVI Ly$\alpha$ and Fe XXV K$\alpha$, is then calculated using the theoretical line emissivity from ATOMDB and assuming a solar abundance.     
The predicted Fe line luminosities as a function of cloud density are shown in Extended Data Figure~7. 
The maximum luminosity ever reached is $3.8\times10^{36}\rm~erg~s^{-1}$ for Fe XXVI Ly$\alpha$ and $1.5\times10^{37}\rm~erg~s^{-1}$ for the sum of the Fe XXV K$\alpha$ triplet. 
We note that these are in broad agreement with the results of ref.\cite{2002A&A...387...76B}, which studied highly ionized Fe lines generated in Compton-thin photoionized gas around a central AGN using a similar numerical approach.
The predicted Fe line luminosities are substantially lower than the observed values of $\sim$$1.9\times10^{38}\rm~erg~s^{-1}$ and $\sim$$1.1\times10^{38}\rm~erg~s^{-1}$ for Fe XXVI Ly$\alpha$ and Fe XXV K$\alpha$, respectively. 
We conclude that photoionization by the central LLAGN alone cannot produce the observed highly ionized Fe lines.

\noindent{\bf The Fe XXV/XXVI lines are not produced in the accretion inflow}\\
Within the truncated radius of $3\times 10^3~r_{\rm g}$, the  
standard thin disk is replaced by a hot accretion flow\cite{2018MNRAS.476.5698Y}, which is sufficiently hot (if not too hot) to produce highly ionized Fe lines.
To investigate this possibility, we have performed 3-dimensional general relativity magnetohydrodynamic (GRMHD) numerical simulations of hot accretion flow onto a black hole of $7\times10^{7}\rm~M_\odot$, following the procedure descibed in ref.\cite{2015ApJ...804..101Y}. Technically it is very hard for the simulated accretion inflow to reach a steady state from the black hole horizon up to 3000 $r_{\rm g}$. Instead, we first simulate the inner region of the accretion flow and obtain the steady state solution from the horizon to 100 $r_{\rm g}$. For this aim, we use the code Athena++
\cite{2016ApJS..225...22W} to solve the ideal GRMHD equations in the Kerr metric. We find that away from the innermost region, all physical quantities such as density, temperature, and velocity can be well fitted by some functions of coordinates such as radius. So we are able to extrapolate the simulation data based on these scaling laws from 100 $r_{\rm g}$ to $r_{\rm tr}=3000~r_{\rm g}$. 

The simulation data are then supplied to the {\it apec} model to calculate the emission lines and the bremsstrahlung radiation.
We find that the plasma temperature remains $\gtrsim8\times10^9\rm~K$ within $\sim800~r_{\rm g}$, such that the gas is fully ionized and hardly radiates any emission lines. The temperature drops gradually to $\sim10^9\rm~K$ between $1000-3000~r_{\rm g}$, where bulk of the Fe XXVI line is produced.
The velocity information is used to calculate the Doppler effect, for a given viewing angle (the angle between the jet axis and the line-of-sight).

The synthetic X-ray spectrum between 6.5--7.3 keV for a viewing angle of 45$^\circ$ is shown in Extended Data Figure 8.
This angle is chosen to mimic the
blueshifted and redshifted components of the observed Fe XXVI Ly$\alpha$ line (A viewing angle of 15$^\circ$, which is more consistent with the jet axis, will cause the two components to overlap, inconsistent with the observed profile.).
While the predicted Doppler shifts of the lines are acceptable, the temperature is too high so the gas is too strongly ionized, hence the predicted equivalent width of the Fe XXVI Ly$\alpha$ line is only $15\rm~eV$, much smaller than the observed value of $\sim$110 eV. 
In principle, a more extended hot accretion flow (i.e., with a larger $r_{\rm tr}$) would cover lower temperatures ($T \propto r^{-1}$) and thus produce stronger Fe XXVI Ly$\alpha$. We have tested such a case and found that, due to the $v _\phi \propto r^{-\frac{1}{2}}$ dependency, the part of the accretion flow that contributes most Fe XXVI Ly$\alpha$ flux will have a much smaller Doppler shift, and the resultant line profile will be centrally peaked or flat-topped, again inconsistent with the observed double-peaked profile. 
Therefore, we conclude that the hot accretion flow itself cannot explain the observed highly-ionized Fe lines. 

We also note that the predicted 2–10 keV luminosity by the accretion inflow from its innermost region to $3000~r_{\rm g}$ is only
$\sim 5\%$ of the observed value, when the gas density is normalized to have an inflow rate of $\sim 2\times10^{-3}\rm\ M_{\odot}\ yr^{-1}$ at $3000~r_{\rm g}$, to be consistent with the outflow rate of the wind. Such a small contribution is consistent with the theory of hot accretion flow, since for such an low-luminosity system of M81*, the jet may dominate the X-ray radiation\cite{2014ARA&A..52..529Y}.

\noindent{\bf Wind simulation and synthetic wind spectrum}\\
Both the hot accretion flow and the truncated thin disk can launch winds, but here we only need to focus on the wind from the hot accretion flow since the temperature of the wind from the thin disk would be too low to be responsible for the observed X-ray spectrum.
Ideally we should simulate the accretion flow and winds simultaneously from the black hole event horizon to the largest radius where the observed iron lines can be emitted. But simple estimation indicates that this largest radius is $\gtrsim 10^5~r_{\rm g}$. Again, it is technically very hard to simulate such a large radial dynamical range. Therefore, we only choose our simulation domain to cover the region where the observed iron lines can be produced. Specifically, the inner and outer boundaries of our simulation are set to be $3000~r_{\rm g}$ and $10^6~r_{\rm g}$, respectively. There is no hot accretion inflow in this region so we only simulate the wind. 

In such a setup, the most important thing is to determine the inner boundary conditions of the wind, i.e., the physical properties of the wind at $3000~r_{\rm g}$. These boundary conditions are extrapolated from the small-scale GRMHD simulation of black hole hot accretion flows by Ref.\cite{2015ApJ...804..101Y,Yang2021}. 
Ref.\cite{2015ApJ...804..101Y} deals with SANE (standard and normal evolution) and a non-spinning black hole, while ref.\cite{Yang2021} extends ref.\cite{2015ApJ...804..101Y} by considering both SANE and MAD (magnetically arrested disk) around black holes with different spins. Using the ``virtual particle trajectory'' approach, these two works have  obtained precisely the properties of the wind, including spatial distributions of rotational and radial velocities, temperature, magnetic field, and density.  
 Although the accretion flows in these two works can achieve steady state only within $\sim 100~r_{\rm g}\ll3000~r_{\rm g}$, it is found that these quantities can be well described by power-law functions of radius, which allows us to reasonably extrapolate the physical quantities to much larger radii such as $3000~ r_{\rm g}$. The result is then used as the inner boundary condition of our wind simulation.

The scaling properties of the wind are as follows. The temperature is $\sim 0.5 T_{\mathrm{vir}}(r) \sim 0.5 GM_{\rm BH}/(rk)$. The radial velocity is $v_r\sim(0.2-0.6)v_\mathrm{k} (r) = (0.2-0.6)(GM_{\rm BH}/r)^{0.5}$, which depends on the spin of the black hole and the accretion mode (SANE or MAD). Since a jet is observed in M81*, we assume that the spin of the central black hole is rapid and the accretion is in the MAD mode\cite{2014ARA&A..52..529Y}. In this case, $v_r\approx 0.6v_\mathrm{k} (r)$. The rotational velocity of the wind is $\sim(0.8-1)v_\mathrm{k} (r)$, and we adopt $v_{\phi}\approx 0.8v_\mathrm{k} (r)$. The thermal-to-magnetic pressure ratio, $\beta \equiv P_{\mathrm{gas}}/P_{\mathrm{mag}}$, is $\sim0.1-0.4$, and we adopt $\beta\approx 0.2$. 
At large radii, the toroidal magnetic field is several times stronger than the poloidal magnetic field, thus we safely ignore the latter.   
Since the wind can push the surrounding ISM away from the region of interest, we simply neglect the ISM, setting a density floor of $10^{-7}{\rm~cm^{-3}}$ in the simulation.

We perform a 2.5-D MHD wind simulation with the ZEUS-MP/2 code using axisymmetric spherical coordinates ($r, \theta$). The rotation of the wind ($v_{\phi}$) is considered. A black hole with $7\times 10^7\rm~M_{\odot}$ is set at the center. 
The simulation grid has a total of $140\times 18$ zones, logarithmically spaced in the radial direction and uniformly spaced in the latitudinal direction. We find that such a resolution is enough since there is little turbulence in the wind. 
The grid in the latitudinal direction extends from $\theta=5^{\circ}$ to $50^{\circ}$. The rotation axis is avoided given the singularity of spherical coordinate and the existence of the jet; the equatorial plane is avoided because the wind only occupies this region when it propagates outward from the hot accretion flow, according to Ref.\cite{2015ApJ...804..101Y} (refer to Figure 1 therein). 
In fact, our current simulation again confirms this result. We also neglect the jet, because the existence of the jet does not affect the dynamics of the wind, and the jet does not produce emission lines. 
At the inner radial boundary we inject the wind following the aforementioned inner boundary conditions. 
As usual, the outflow boundary condition is adopted at $10^6 ~r_{\rm g}$, the axisymmetric boundary condition is adopted at $\theta = 5^{\circ}$ , and the reflecting boundary condition is adopted at $\theta = 50^{\circ}$.

The 2-D distributions of temperature and density of the simulated wind are shown in Figure~2b\&c, overlapped with the velocity vector. 
The 2-D grid is then expanded into a 3-D grid by assuming axisymmetry about the jet axis and mirror symmetry about the equatorial plane.
We are now in a place to predict the wind X-ray spectrum.
To do so, we first calculate the density-weighted thermal emission for each grid in the black hole rest frame, again utilizing ATOMDB and assuming solar abundance.   
The synthetic spectrum is then produced by integrating along a given viewing angle. Doppler effect is included according to the projected velocity in each grid. 
Gravitational lensing effect is also taken into account, which 
slightly boost the redshifted component (i.e., from behind the SMBH in the case of an outflow) and partially compensate for the Doppler dimming effect. 
The absolute normalization of the gas  density is determined by matching the integrated spectrum to the observed spectrum (Figure~1d).
We have verified that self-absorption in the Fe XXVI Ly$\alpha$ and XXV K$\alpha$ lines is negligible for an equivalent column density $\lesssim10^{22}\rm~cm^{-2}$, which applies to the wind as well as the hot accretion flow.

\end{methods}

\begin{addendum}
\item[Acknowledgements]
This research made use of observations taken from the {\it Chandra} X-ray Observatory, software package CIAO provided by the {\it Chandra} X-ray Center and spectral fitting package XSPEC,
and the High Performance Computing Resource in the Core Facility for Advanced Research Computing at Shanghai Astronomical Observatory. 
Z.L. and F.S. acknowledges support by the National Key Research and Development Program of China (2017YFA0402703) and Natural Science Foundation of China (grant 11873028).
F.Y. and B.Z. are supported in part by the National Key Research and Development Program of China (Grant No. 2016YFA0400704), the Natural Science Foundation of China (grant 11633006), and
the Key Research Program of Frontier Sciences of CAS (No. QYZDJSSW-SYS008). We wish to thank Can Cui and Hai Yang for help with initial model tests and GRMHD simulation of hot accretion flows, and Junfeng Wang for helpful discussions.

\item[Author contributions]
This research program was designed and framed by Z.L. and F.Y. Analysis and modelling of the X-ray data were performed by F.Z. with the help of Z.L. Numerical simulations of the accretion flow and wind were performed by B.Z. and F.Y. 
All authors were involved in the discussion and interpretation of the results presented, and all contributed to writing the paper.

\item[Competing interests] 
The authors declare that they have no competing financial interests.  

\item[Additional information] 
Correspondence and requests for materials should be addressed to Z.L. (email: lizy@nju.edu.cn) or F.Y. (email: fyuan@shao.ac.cn).

\end{addendum}
\clearpage

\begin{figure}
\centering\includegraphics{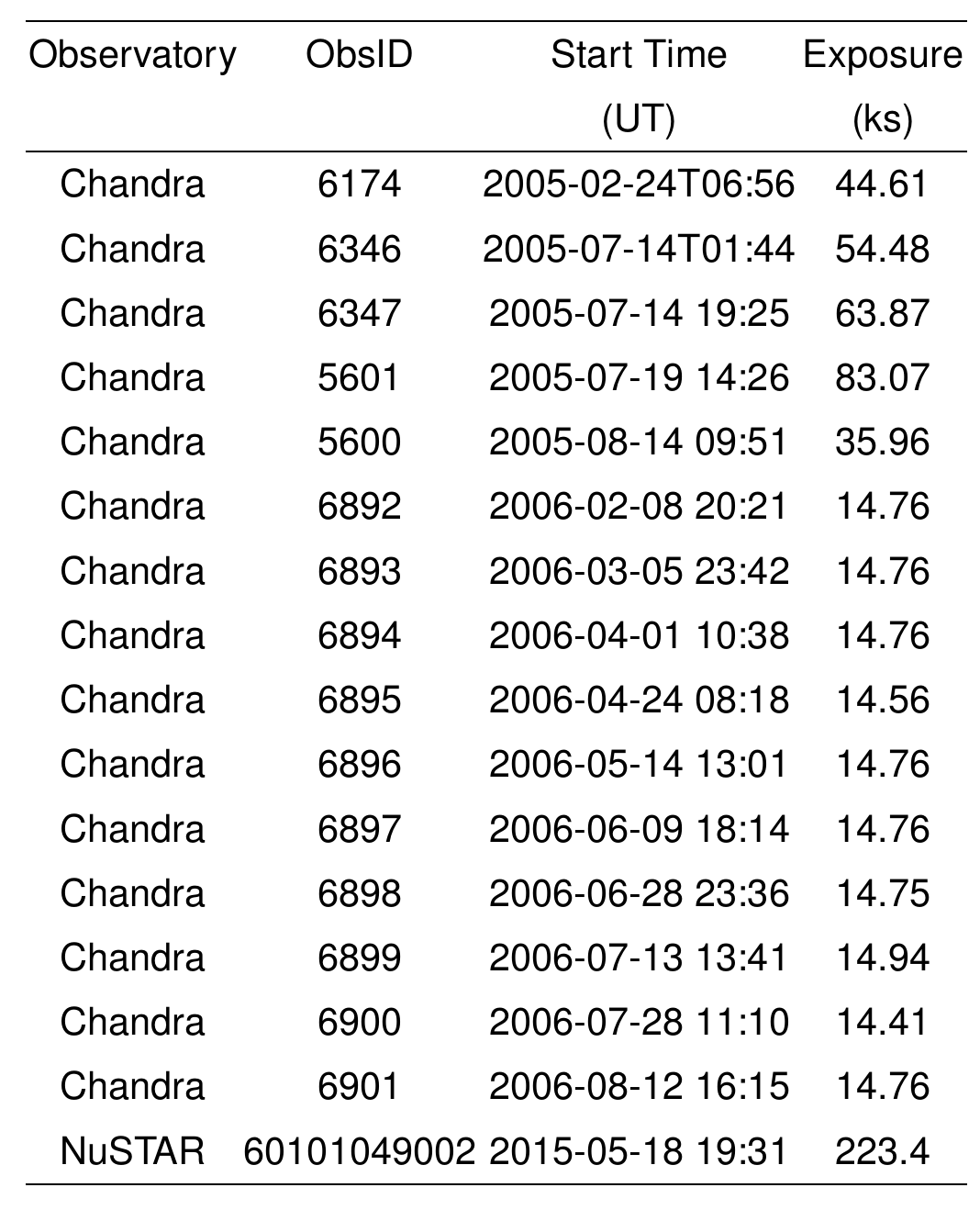}

\caption*{{\bf Extended Data Figure 1. Log of X-ray observations.}}
\label{fig:obs_log}
\end{figure}
\clearpage

\begin{figure}
\centering\includegraphics[width=0.8\textwidth,angle=0]{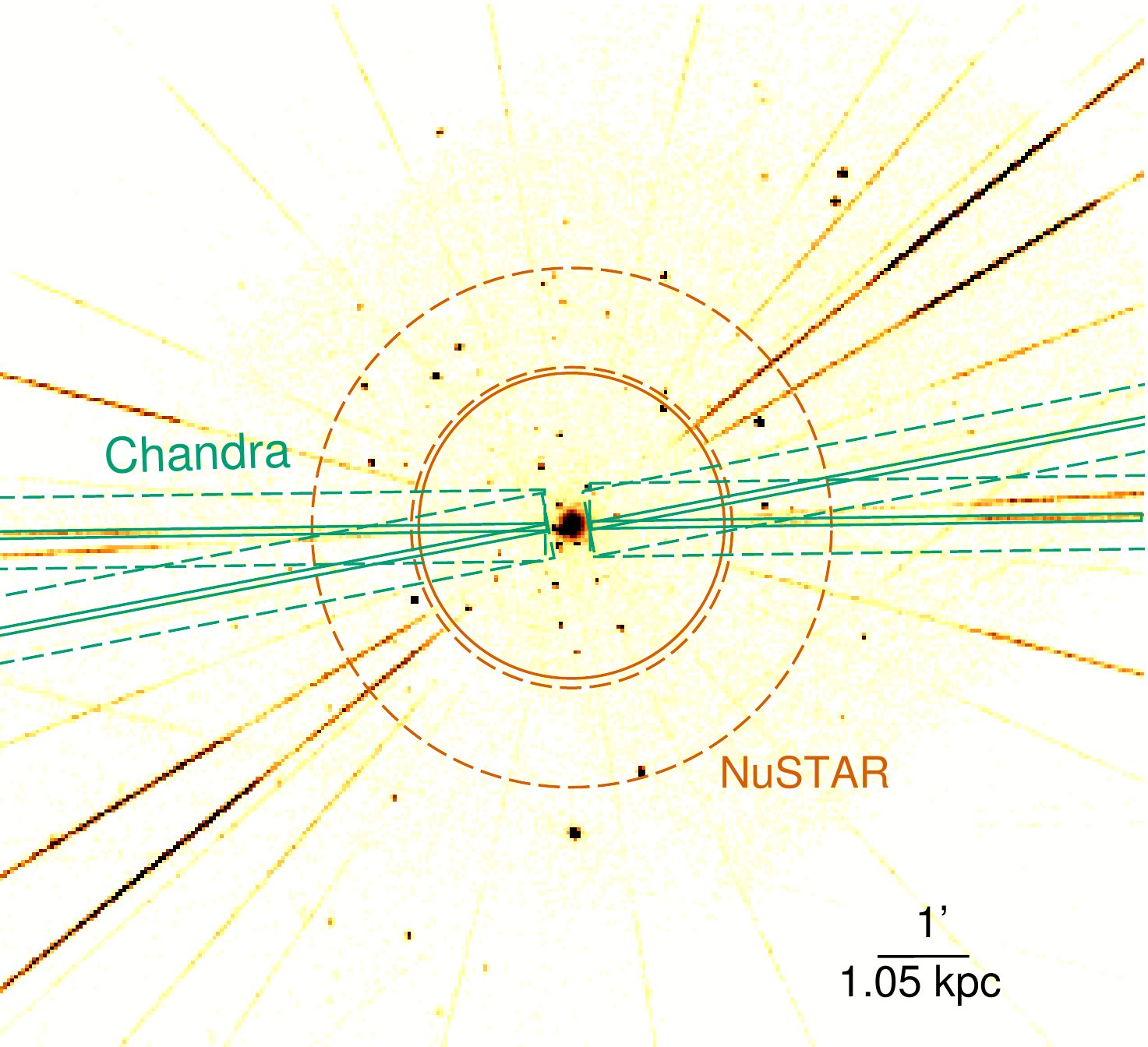}

\caption*{{\bf Extended Data Figure 2. The combined {\it Chandra}/HETG images highlighting the 1st-order MEG/HEG arms.} 
Spectra are extracted from the individual observations of different roll angles and then combined to form the final spectrum. 
The green rectangles illustrate the spectral extraction region (solid rectangles for the source and the adjacent dashed rectangles for the background) for one of the 15 observations. 
The {\it NuSTAR} source region is marked by the vermillion solid circle, while the corresponding background region is marked by the vermillion dashed annulus. Discrete sources falling within the {\it NuSTAR} spectral extraction regions are not excluded, but their collective flux contribution is negligible.}
\label{fig:FoV}
\end{figure}

\begin{figure}
\centering
\includegraphics[width=0.7\textwidth,angle=270]{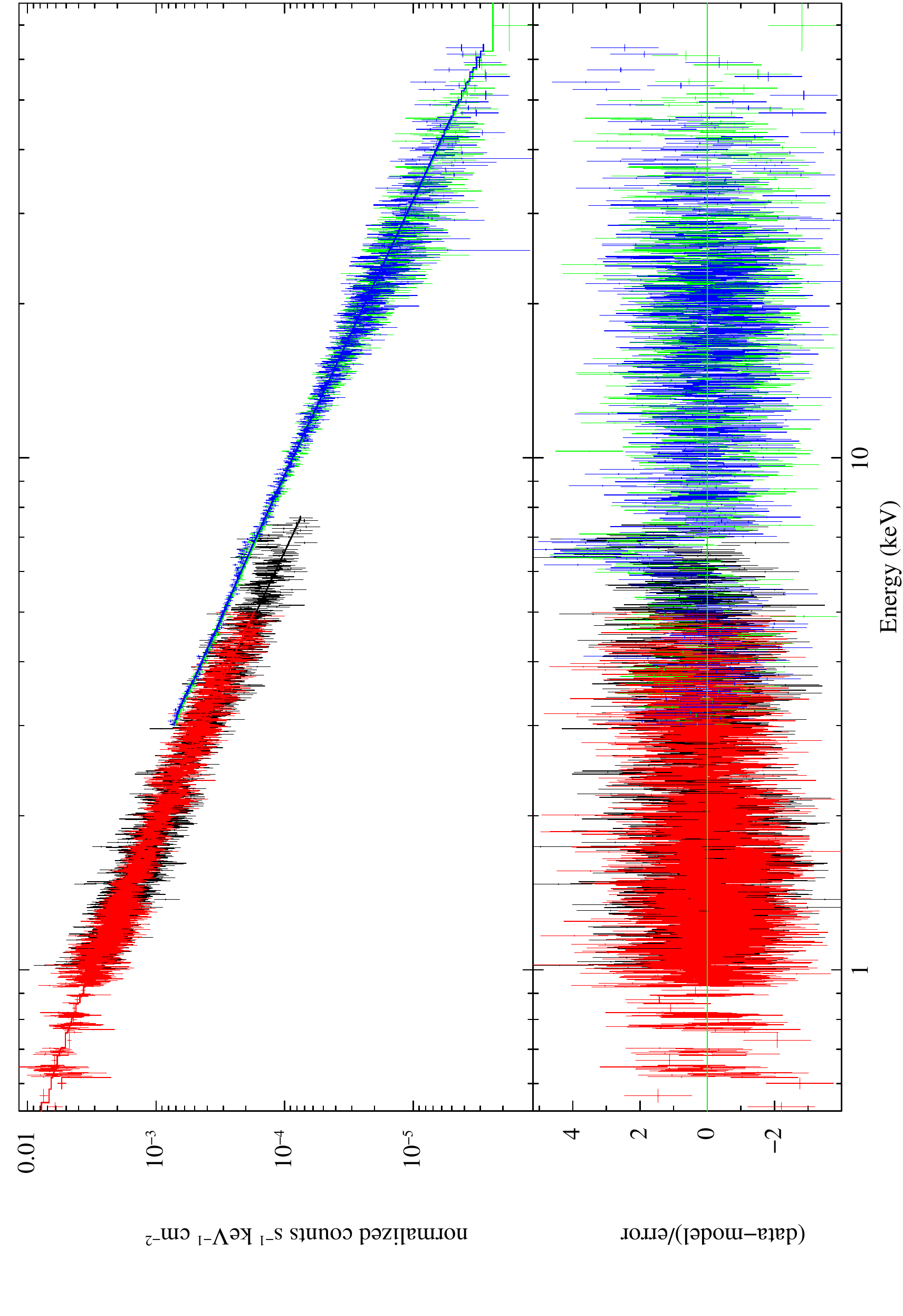}

\caption*{{\bf Extended Data Figure 3. Observed X-ray spectra of M81*.} Red: coadded {\it Chandra}/MEG 1st-order spectrum; Black: coadded {\it Chandra}/HEG 1st-order spectrum;
Green: {\it NuSTAR}/FPMA spectrum;
Blue: {\it NuSTAR}/FPMB spectrum.
The error bars are of 1$\sigma$.
The best-fit absorbed power-law model is shown by the solid lines. 
The ratio of residual/error is shown in the bottom panel.
Significant excess is seen between 6--7 keV due to the presence of Fe lines. 
The spectra shown here are binned to achieve a S/N greater than 3 for better illustration, while in the actual spectral fit throughout this work the spectra are binned to have at least one count per bin to optimize the spectral resolution.}
\label{fig:baseline}
\end{figure}

\begin{figure}
\centering
\includegraphics[width=1\textwidth,angle=0]{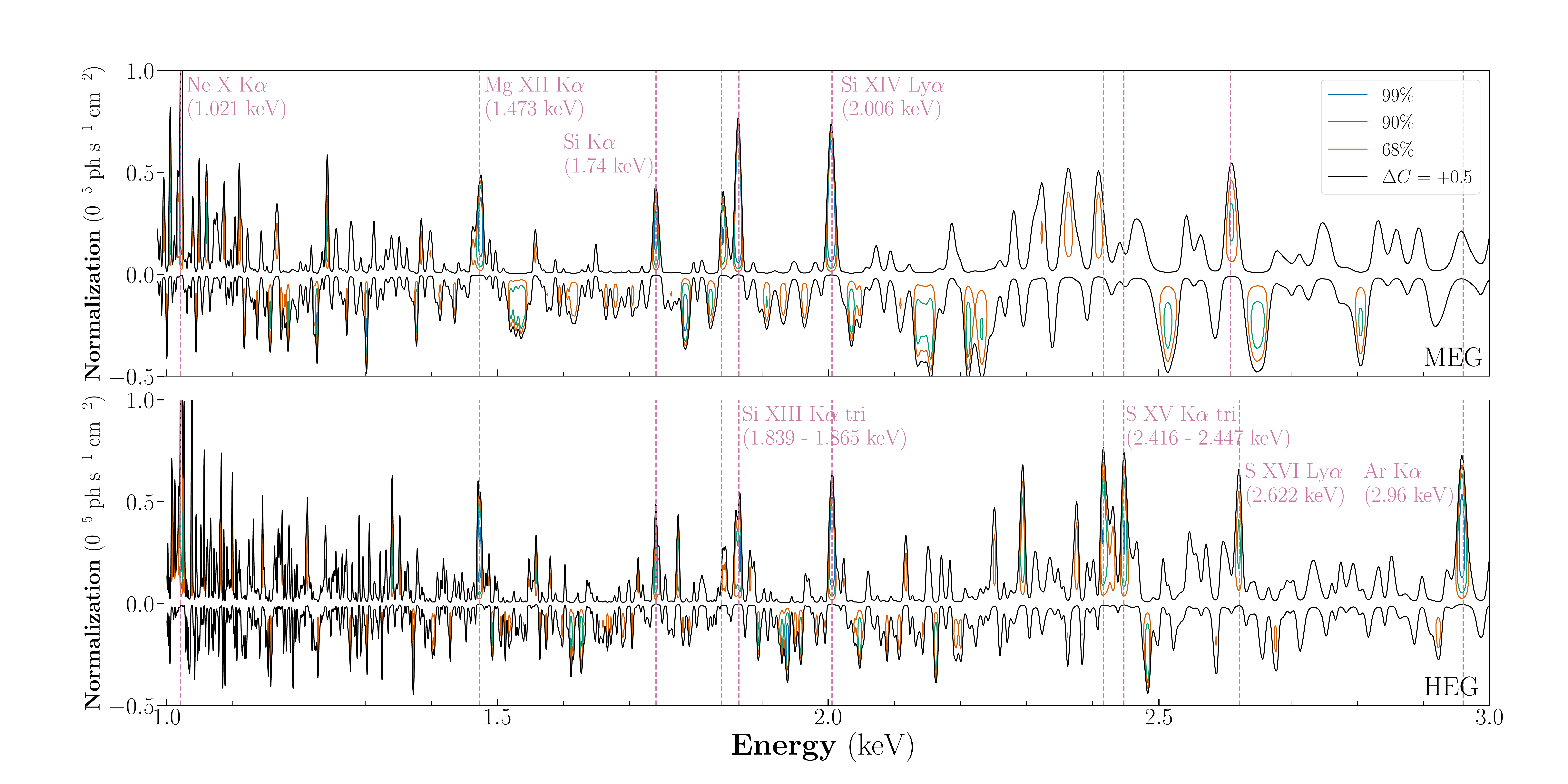}

\caption*{{\bf Extended Data Figure 4. Blind line search of the MEG and HEG spectra over 1--3 keV.} Blue, green and orange contours indicate confidence level of 99\%, 90\% and 68\%,  respectively, of a test line according to the differential $C$-stat value against the baseline continuum model. The black contour denotes where $\Delta C=+0.5$. Significant lines with an identified atomic transition are denoted with the pink vertical dashed lines.}
\label{fig:blind_search_1}
\end{figure}

\begin{figure}
\centering\includegraphics{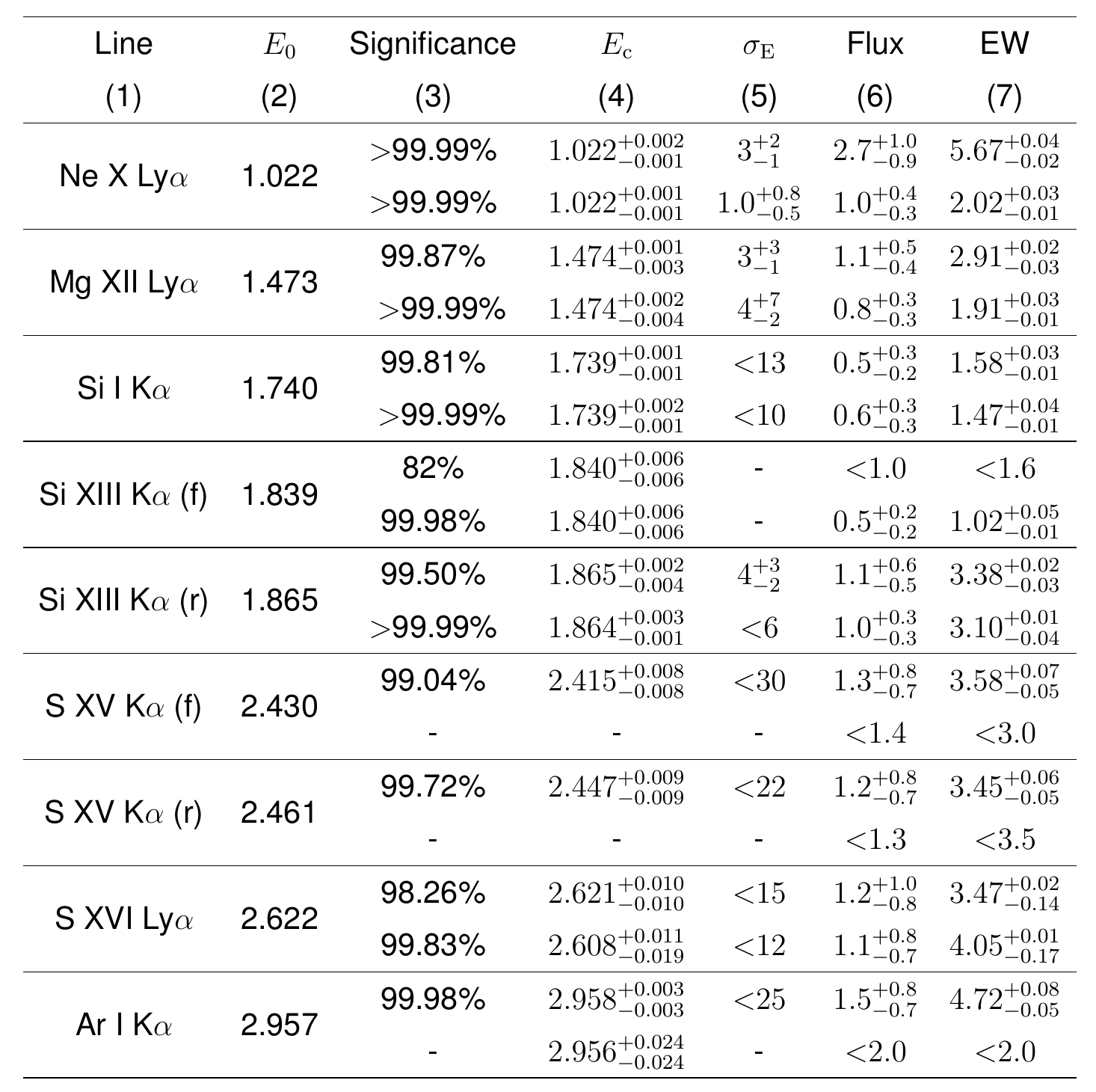}

\caption*{{\bf Extended Data Figure 5. Additional Emission lines in the HETG spectrum.} 
(1) Each identified transition is followed by two rows, the upper row for the HEG measurement and the lower row for the MEG measurement. (2) Rest-frame energy of the identified transition, in units of keV. (3) Significance of the line. (4) The best-fit central energy in units of keV. (5) The Gaussian line width, in units of eV. $3\,\sigma$ upper limit is provided for unresolved lines. (6) Line flux in units of $\rm10^{-14}~erg~cm^{-2}~s^{-1}$. 
$3\,\sigma$ upper limit is provided for undetected lines.
(7) Equivalent width in units of eV. Quoted errors are at 90\% confidence level.}
\label{fig:line_add}
\end{figure}
\clearpage

\begin{figure}
\centering\includegraphics[width=1\textwidth,angle=0]{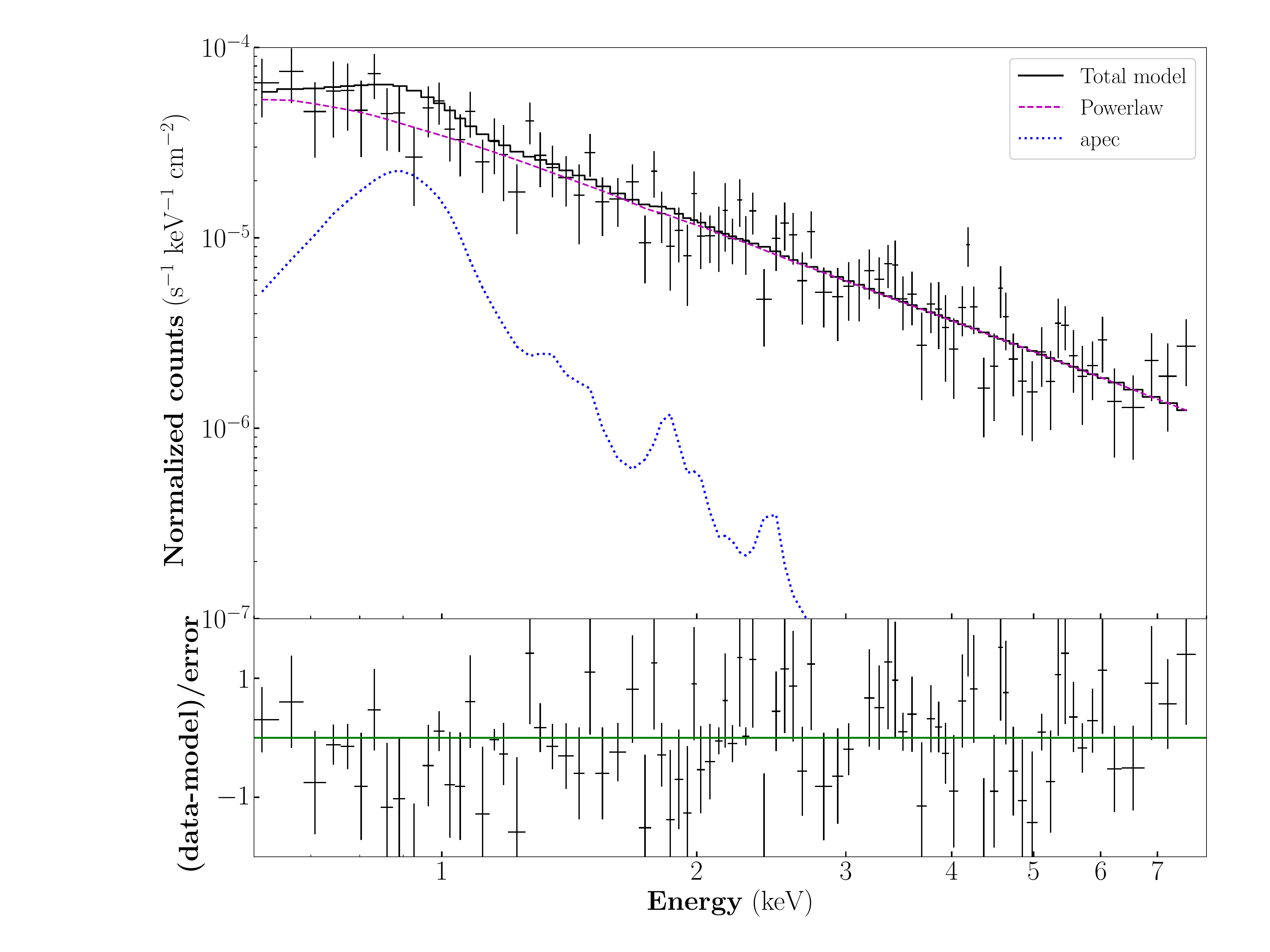}

\caption*{{\bf Extended Data Figure 6. Coadded HETG zeroth-order spectrum from an annulus of inner-to-outer radii of 2.5$''$ --5.0$''$ around M81*.} The spectrum can be well fitted by an absorbed power-law, shown as the magenta line. 
The error bars are of 1$\sigma$. The ratio of residual/error is shown in the bottom panel.
An additional thermal component, represent by an {\it apec} model with a plasma temperature of 0.9 keV, is allowed by the data and shown as the blue line. 
The sum of the power-law and {\it apec} is plotted as the black line.}
\label{fig:offnucleus}
\end{figure}

\begin{figure}
\centering\includegraphics[width=0.8\textwidth,angle=0]{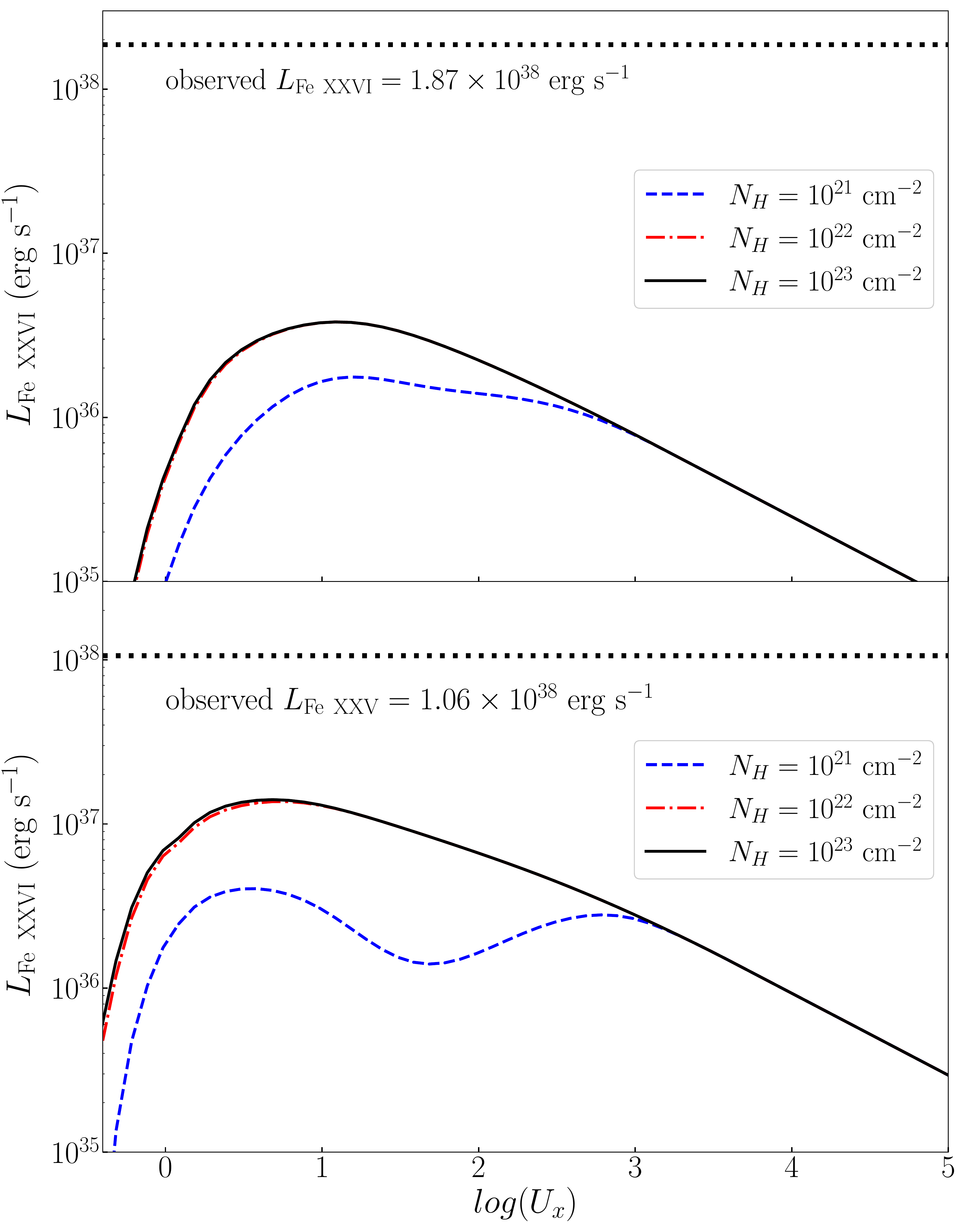}

\caption*{{\bf Extended Data Figure 7. Predicted Fe line luminosity of an isotropic and uniform gas cloud photonionized by a central AGN.} 
The upper (lower) panel is for Fe XXVI Ly$\alpha$ (XXV K$\alpha$). The ionization parameter is evaluated for an intrinsic X-ray spectrum same as M81* and over photon energy of 2--10 keV. 
The cloud has an equivalent hydrogen column density of $N_{\rm H}=10^{23}{\rm~cm^{-2}}$ (black solid line), $10^{22}{\rm ~cm^{-2}}$ (red dash-dotted line) and $10^{21}{\rm~cm^{-2}}$ (blue dashed line). 
The black dotted horizontal line in each panel marks the observed line luminosity, which is substantially higher than the predicted values.}
\label{fig:L_Fe}
\end{figure}

\begin{figure}
    \centering
    \includegraphics[width=1\textwidth,angle=0]{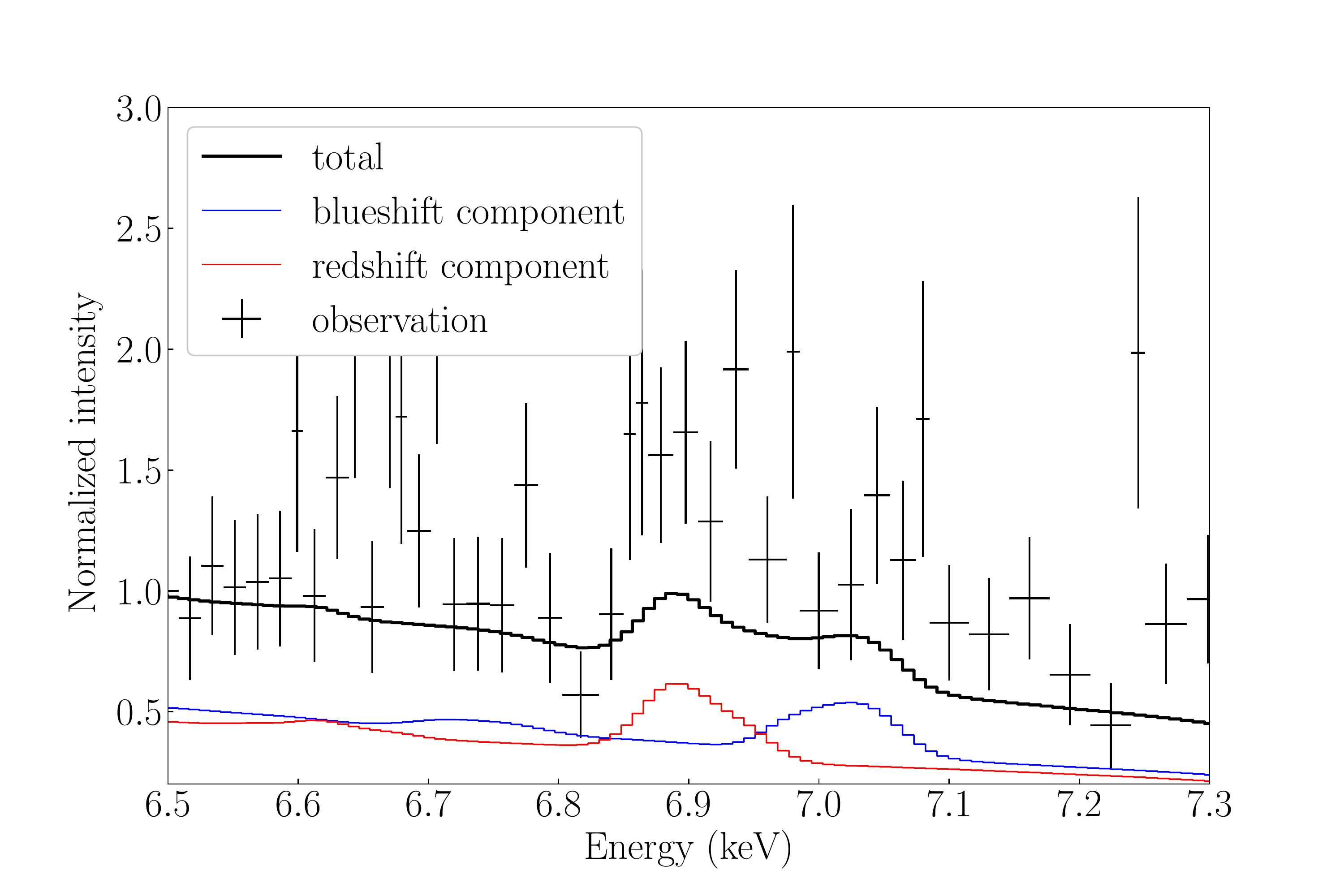}
    
\caption*{{\bf Extended Data Figure 8. Predicted 6.5--7.3 keV spectrum from simulation of the hot accretion flow.} The viewing angle is set to be $45^\circ$ with respect to the jet axis. The blue, red and black curves show the blueshifted, redshifted and total spectrum, respectively. The spectra have been convolved with the HEG instrumental response. 
The black crosses mark the observed spectrum as a reference. The error bars are of 1$\sigma$.
The observed Fe XXVI and XXV lines have an equivalent width substantially higher than predicted by the hot accretion flow.}
    \label{fig:ADAF_lp}
\end{figure}
\clearpage

\end{document}